\documentclass[global,referee,twocolumn]{svjour}

\usepackage{graphicx}
\usepackage{dcolumn}
\usepackage{bm}
\usepackage{color}
\usepackage[normalem]{ulem}

\begin{document}

\title{Off-axis retrieval of orbital angular momentum of light  stored in cold atoms}

\author{R.~A.~de Oliveira$^{1, \star}$, L.~ Pruvost$^{2}$, P. ~S.~Barbosa$^{1}$, W.~S.~ Martins$^{1}$,  S.~Barreiro$^{1, \ast}$, D.~Felinto$^{1}$, D.~ Bloch$^{3}$, and J.~W.~R.~Tabosa$^{1}$}
 \institute{$^{1}$ Departamento de F\'{\i}sica, Universidade Federal de Pernambuco, 50670-901 Recife, PE - Brazil\\
$^{2}$ Laboratoire Aim\'{e} Cotton, CNRS, Univ Paris-Sud, ENS Cachan, 91405 Orsay, France\\$^{3}$ Laboratoire de Physique des Lasers, CNRS, Universit\'{e} Paris 13-Sorbonne Paris Cit\'e, 93430 Villetaneuse, France}

\date{\today}
\authorrunning{R.~A.~de Oliveira \textit{et al.}}
\titlerunning{Off-axis retrieval of orbital angular momentum of light...}
\maketitle
\begin{abstract}
We report on the storage of orbital angular momentum (OAM) of light of a Laguerre-Gaussian mode in an ensemble of cold cesium atoms and its retrieval along an axis different from the incident light beam. We employed a time-delayed four-wave mixing configuration to demonstrate that at small angle ($2^{o}$), after storage, the retrieved beam carries the same OAM as the one encoded in the input beam. A calculation based on mode decomposition of the retrieved beam over the Laguerre-Gaussian basis is in agreement with the experimental observations done at small angle values. However, the calculation shows that the OAM retrieving would get lost at larger angles, reducing the fidelity of such storing-retrieving process. In addition, we have also observed that by applying an external magnetic field to the atomic ensemble the retrieved OAM presents Larmor oscillations, demonstrating the possibility of its manipulation and off-axis retrieval.
\end{abstract}

\section{Introduction}
\indent Light beams carrying orbital angular momentum have attracted a great interest in recent years owing to several possibilities of applications, ranging from the mechanical manipulation of macroscopic particles to the encoding of quantum information \cite{Allen04,Torner13}. The direct observation of the transfer of quantized OAM of light to a Bose-Einstein condensate was also demonstrated in \cite{Andersen06}. A family of these light beams is constituted by Laguerre-Gaussian (LG) modes. They are specified by an integer number $\ell$, given the topological charge, associated to the helicoidal phase structure and corresponding  to an OAM per photon equal to $\ell\hbar$ \cite{Allen92} . LG modes constitute a basis and are used to encode information, each LG mode being a bit. It was predicted in \cite{Bechmann-Pasquinucci00} that for quantum information and computation purposes the use of multidimensional state space, as the one spanned by OAM of LG modes, leads to a more efficient quantum processing. Therefore, the capability to store, manipulate and retrieve different OAM quantum states is of crucial  importance. 

Similarly to other states of light, the quantum information encoded in modes with OAM can be stored in an atomic ensemble through the light-atom interaction. The nonlinear interaction of LG beams with atomic systems has been investigated via four-wave mixing processes \cite{Petrov99} and the corresponding conservation of OAM demonstrated both in cold and thermal atoms  \cite{Barreiro03,Guo06}. The storage of OAM in cold and thermal atomic ensembles was also previously demonstrated, both in the classical \cite{Pugatch07,Moretti09} and single photon \cite{Dong13,Nicolas14} regimes. In the context of  quantum computing, the generation of twin light beams carrying OAM with quantum intensity correlation have been demonstrated through four wave mixing (FWM) in thermal rubidium vapor \cite{Marino08} and also the entanglement of the OAM of photon pairs \cite{Kozuma06,Guo08}. OAM storage/retrieval is in a sense similar to images storage/retrieval recently investigated with FWM process \cite{Wu13}.

The experiments previously done on OAM storage and retrieval mainly used a collinear configuration where the retrieving beam was collinear to the incident OAM-encoded input. With this geometry, the conservation of OAM was a natural deduction. Retrieving OAM in a  direction which is different from the OAM-encoded input one is an important issue associated with quantum memories for OAM states not only concerning the open question of conservation of OAM in such case, but also the possibility to separate the writing axis from the retrieving one, as well as to distribute OAM-encoded information along different directions in space. 

In this work we investigate, both experimentally and theoretically, the off-axis retrieval of OAM stored in cold cesium atoms. Our experimental configuration makes use of the holographic interpretation of the FWM process  with two grating writing beams corresponding to an OAM-encoded beam and a gaussian beam and a time delayed gaussian reading beam generating the OAM retrieval along a direction different from the OAM-encoded input beam. The retrieving direction makes a small angle with the input one.  We experimentally show that, after a storage time of some micro-seconds, the OAM is conserved in the case of a small angle . We also demonstrate that a superposition of an OAM mode ($\ell\neq0$) with a gaussian one ($\ell=0$)  is also stored/retrieved with a good  fidelity. The OAM conservation law holds for angles of order of some degrees, therefore allowing the spatial separation of the two beams (encoded and retrieved ones), which can be of considerable practical importance. It is worth noticing that our experiment differs from the one in \cite{Guo06} because even if a non-collinear transfer of OAM was observed, there was no storage of OAM and the experiment was made more complex by a non-degenerate multi lasers configuration, with several OAM-carrying beams as inputs in the cold atom sample.

 Performing a mode decomposition of the retrieved beam (tilted OAM beam) we have studied the effect of the angle value on the purity of the retrieved mode. The calculations show that at angles larger than some degrees the purity is degraded. It gives a limitation of the off-axis retrieval method and also shows that the results obtained in  \cite{Guo06} can not be generalized for arbitrary values of angles. The  conservation of OAM had also been analyzed in spontaneous parametric down conversion where the pump, signal and idler beams were not collinear \cite{Molina-Terriza03,Torres04}. In this different situation the analysis had shown that although the OAM conservation law is verified for small angles between the pump beam and the signal and idler beams, the law is violated when the generated beams make large angles with the pump beam. 

Finally we report on the observation of Larmor oscillation of the stored OAM in the presence of an external magnetic field, which also demonstrates that the retrieval of the stored OAM along a different direction is possible after its manipulation.

\section{Experimental results}
\indent The experiment is performed in cold cesium atoms obtained from a MOT. We employ the degenerate two-level system associated with the cesium hyperfine transition $6S_{1/2}(F=3)\leftrightarrow 6P_{3/2}(F^{\prime}=2)$. By properly choosing the polarizations of the incident beams one can excite different sets of $\Lambda$ three-level systems with two degenerate Zeeman ground states and one excited state, as shown in  Figs. 1 (a)-(b).  The atoms are initially pumped into the ground state $6S_{1/2}(F=3)$  via a non resonant excitation induced by the trapping beams, after we switch off  the MOT repumping beam and the quadrupole magnetic field. Our light storage scheme corresponds to a time delayed four-wave mixing configuration where the beams are incident into the medium according to the time sequence shown in Fig. 1(c) . The grating writing beams, $W$ and $W^{\prime}$, correspond respectively to the OAM-encoded LG beam and the gaussian beam. These beams, with the same frequency and opposite circular polarizations, propagate along the directions $z$ and $z^{\prime}$ respectively, and are incident in the cloud forming a small angle $\theta \approx 2^{o}$. These writing beams are kept on for a long period ($\approx 30 \mu s$), so to create a stationary Zeeman ground state coherence grating. After a  storage time $t_{s}$, measured after the turning off of the writing beams, the reading beam $R$, counter-propagating and with opposite circular polarization to the writing beam $W$, is turned on. The retrieved beam $C$, whose propagation direction is opposite to $W^{\prime}$ (as determined by the phase matching condition) is monitored either with a CCD camera or a fast photodiode. Also, during all the measurements three pairs of Helmholtz coils are used to compensate stray magnetic fields.

In the experiment the incident OAM-encoded beam ($W$) is a LG mode, with $\ell=1,2,3,4$, produced by a spatial light modulator (Hamamatsu, LCOS-SLM, Mo. X10468), with resolution 800X600 and 80\% reflectivity efficiency. The different fork phase holograms associated with different values of $\ell$ have approximately the same diffraction efficiency, leading to the same beam power of $ 90 \mu W$. This beam is focused to a diameter much smaller then the atomic cloud diameter which is about $\approx$ 2 mm. The LG ring radius slightly depends on $\ell$ as $\rho_{\ell}=\sqrt{\ell/2}w_{0}$, where $w_{0}=250 \mu m$ is the gaussian beam waist at the center of the MOT, so, for $\ell=1$ to $\ell=4$, the maximum LG ring diameter is of order of $0.7$ mm. The writing beam $W^{\prime}$ has approximately a power of $30 \mu W$ and a spatial gaussian mode with diameter twice larger than that of the $W$ beam. On the other hand, the reading beam $R$ has a power of 1 mW and also a gaussian mode with a diameter slightly larger than the atomic cloud.  All the beams are provided by an ECDL laser and their common frequency is locked to the transition $6S_{1/2}(F=3)\leftrightarrow 6P_{3/2}(F^{\prime}=2)$. Two independent pairs of acousto-optical modulators, working in double passage, allowed us to generate the time sequence depicted in Fig. 1 (c). 

In a first set of experiments, we have measured the topological charge of the incident OAM-encoded beam and the retrieved beam using the method developed in \cite{Singh12}, where the image of a LG mode with topological charge $\ell$, after passing through a astigmatic tilted lens will show a self interference pattern with a Hermite-Gaussian mode shape, presenting $\ell$ minima with the orientation of the pattern $+ 45^{o}$ or $-45^{o}$ depending on the sign of $\ell$. In the column (a) of Fig. 2 we shown the transversal profile of the incident OAM-encoded beams for different values of $\ell$ and in the column (b) the corresponding self-interference patterns which check the input topological charge. In column (c) we show the self-interference patterns of the corresponding retrieved C beams, after a storage time $t_{s}\approx 0.5 \mu s$. As we can see, the incident and retrieved beams have topological charge of same $|\ell|$ value but of opposite signs, which demonstrates that the beams carry the same OAM since they are counter-propagating beams.

\begin{figure}[htb]
  \centerline{\includegraphics[width=8.5cm]{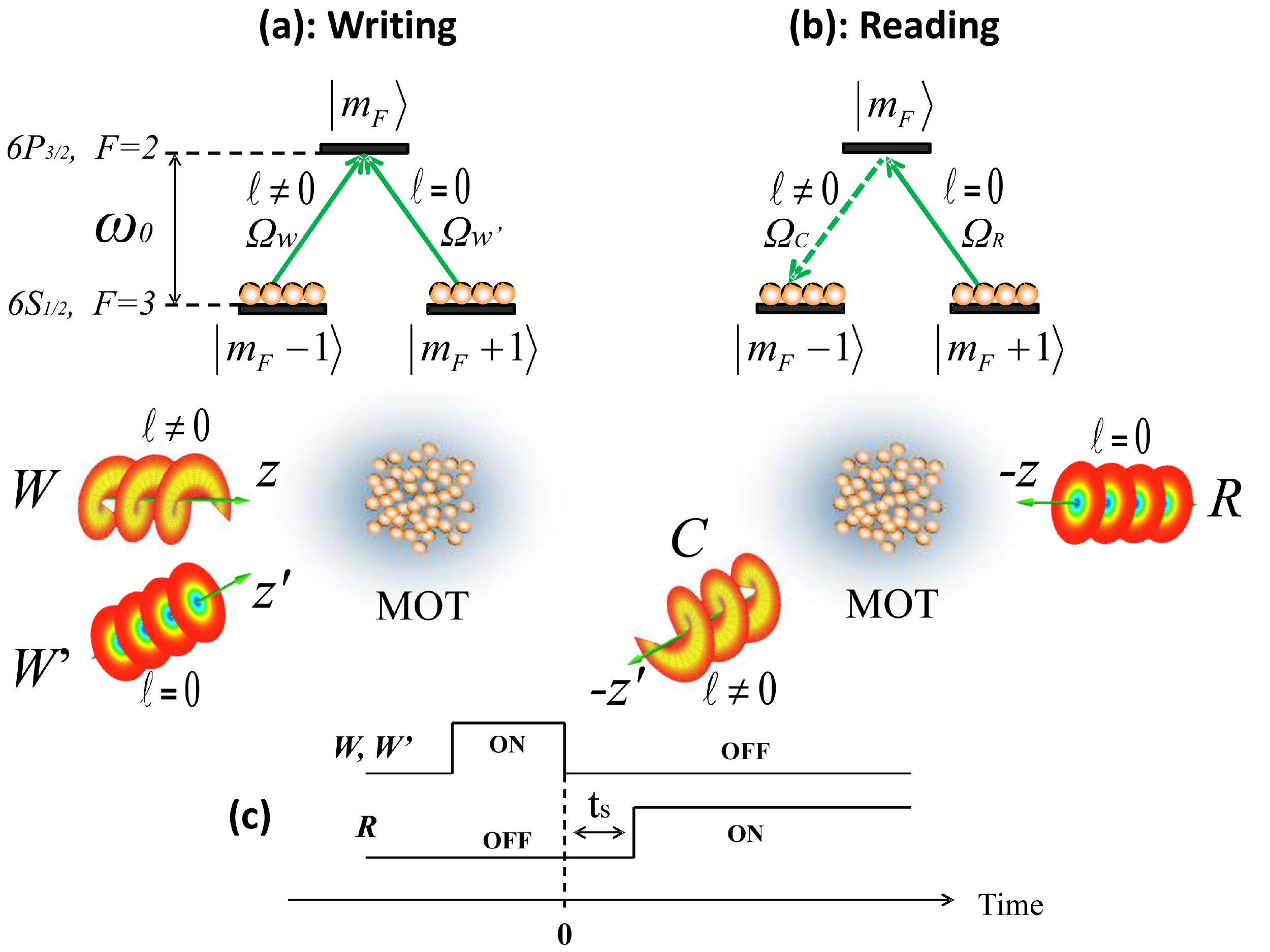}}
  \vspace{-0.5cm}
  \caption{(Color online) A generic Zeeman three-level system,
indicating the coupling of each state with the respective optical fields during the writing (a)
and reading (b) processes.
Propagation directions of the writing beams ($W$, $W^{\prime}$) are along the axis $z$, $z^{\prime}$, respectively, while the reading
($R$) and the generated ($C$) beams counter propagate, respectively, along these directions. (c) Time
sequence for writing ($W$,$W^{\prime}$) and reading
($R$) beams. $t_s$ is the storage time.}
  \label{fig:Fig1}
\end{figure}

\begin{figure}[htb]
  \centerline{\includegraphics[width=7.5cm]{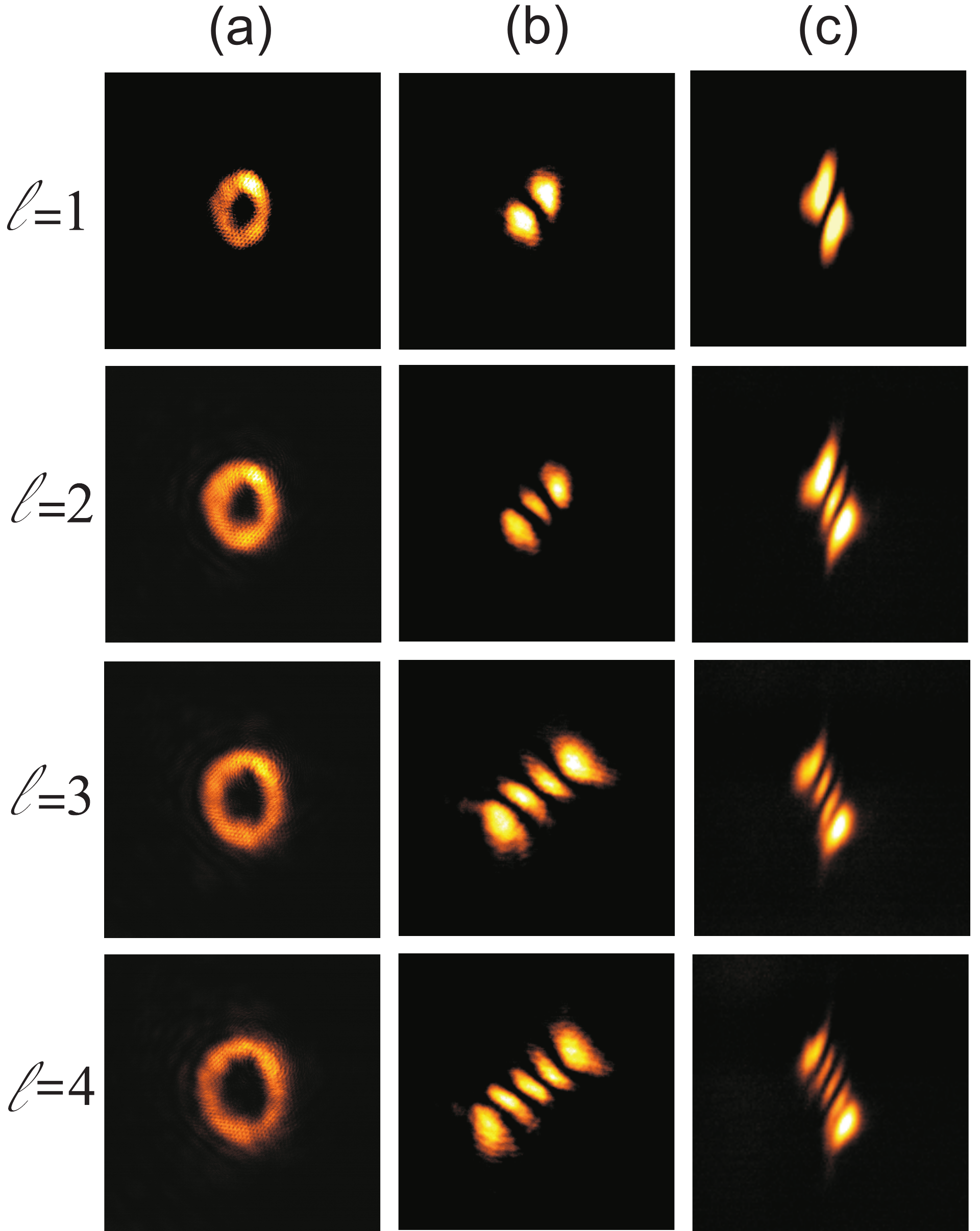}}
  \vspace{-0.5cm}
  \caption{(Color online) (a) Incident beam transverse profile for different values of the topological charge. (b) Corresponding images after passing a tilted lens revealing the topological charge. (c) Retrieved beam images after passing through another tilted lens. }
  \label{fig:Fig2}
\end{figure}

\begin{figure}[htb]
  \centerline{\includegraphics[width=7.0cm, angle=0]{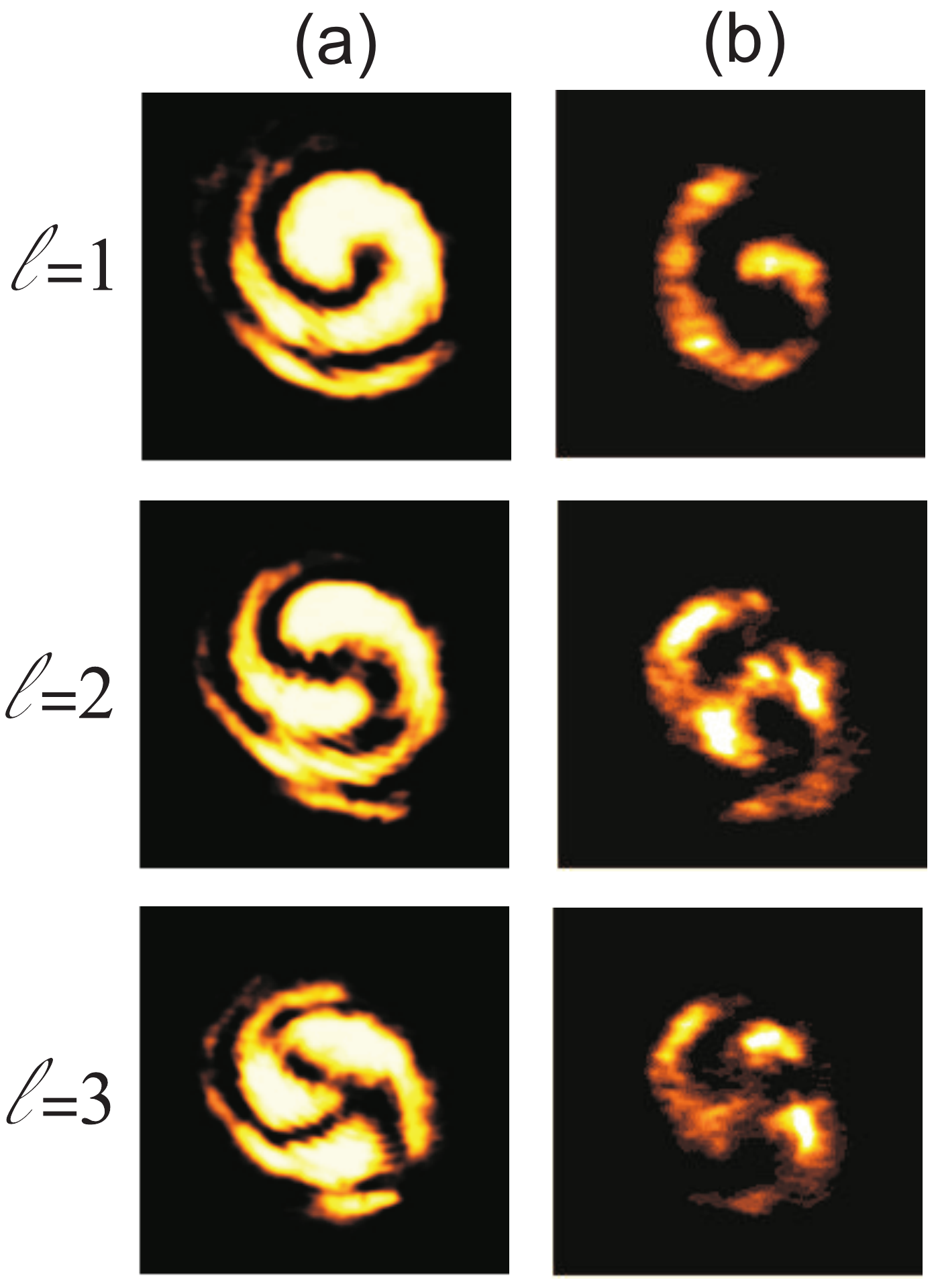}}
  \vspace{-0.2cm}
  \caption{Column (a) shows the interference pattern associated with the incident superposition of LG beams with a gaussian beam, respectively for $\ell=1, 2, 3$. Column (b) shows the corresponding interference patterns retrieved after a storage time of $t_{s}=2 \mu s$,  as described in the text.}
    \label{fig:Fig3}
\end{figure}

We also have demonstrated the storage and non-collinear retrieval of a combination of modes, namely the sum of a LG mode and a gaussian one. For this, we superimposed the incident LG mode with an auxiliary gaussian beam, having a different curvature radius. The so-obtained $W$ beam retro-reflected right before the MOT towards the CCD camera exhibits a spatial intensity distribution shown in column (a) of Fig. 3, which is a classical spiral interference pattern. The number of the branches of the spiral is equal to $\ell$. Column (b) of Fig. 3 shows the corresponding patterns for the retrieved beam in the case of input charges  $\ell=1,2,3$ and after a storage time of $t_{s}=2 \mu s$. By taking into account the change in the sense of the spiral induced by the retro-reflection of the incident superposition, as well as by the relative position of beams foci, we conclude that the OAM of the retrieved beam has the same value and sign as the one of the $W$ beam. This observation is consistent with the previous measurement.

In another series of measurements we have investigated the manipulation of the stored Zeeman coherence grating by the application of an external magnetic field. As demonstrated previously  in \cite{Lezama10}, the application of a dc magnetic field leads to Larmor oscillations of the stored grating whose period is determined by the magnitude of the magnetic field. In the present off-axis retrieval, we have observed Larmor oscillations for $\ell=1, 2, 3 , 4$, with approximately the same period of  $\approx 7 \mu s$ when an external magnetic field of $\approx$ 0.3 Gauss is applied.  Within our experimental uncertainty, the observation of a slight decreasing in the decay time for increasing values of the topological charge was experimentally verified to be associated with the inhomogenities of stray magnetic field. Moreover, similarly to our previous observations in \cite{Moretti09}, we have experimentally observed that the retrieved beam preserves its OAM during the Larmor oscillation, which also demonstrates that the stored OAM can be manipulated and retrieved along a different direction.

\section{Theoretical mode decomposition}

\indent The theoretical analysis is based on a previous work \cite{Moretti08}, which showed that the retrieved field originates from the diffraction of the reading beam into the Zeeman coherence grating and is determined by the optical coherence \cite{Moretti09,Moretti08}, which in the low intensity limit is given by:

\begin{equation}
\sigma_{m_{F},m_{F}-1}(\vec{r},t) \propto g_R(t) e^{-\gamma
t_s}\Omega_R(\vec{r})\Omega_W(\vec{r})\Omega_{W^{\prime}}^*(\vec{r}),
\end{equation}
where, according to Fig. 1 (a) and (b), $\Omega_i(\vec{r})$$ (i=W, W^{\prime}, R)$
represents the Rabi frequency associated with beam $i$ in an atom
at position $\vec{r}$, and is given by
$\Omega_i(\vec{r})=\frac{\mu_{i} {\cal E}_i(\vec{r})e^{i
\vec{k}_i\cdot \vec{r}}}{\hbar}$, with $\mu_{i}$ being the dipole moment
of the transition coupled by $\Omega_i$, ${\cal
E}_i(\vec{r})$ the electric field envelope of beam $i$, and
$\vec{k}_i$ its wave-vector. The term $e^{-\gamma t_s}$ in $Eq. 1$ is a
consequence of the decay of the ground state coherence
due to residual inhomogeneous magnetic field, with $\gamma$ being an effective homogenous
decay rate and $t_s$ the storage time. Finally, the function $g_R(t)$
describes the shape of the retrieved light pulse in mode $C$ and is defined in \cite{Moretti08}.

According to Eq. (1), if $\vec{k}_W = -\vec{k}_R$, the optical coherence generates a field which propagates along $ -\vec{k}_{W^{\prime}}$, consistent with the linear momentum conservation law for the FWM process. Eq. (1) also implies the conservation of the transverse phase. In the situation of our experiment the incident fields $W^{\prime}$ and $R$ are transversally wide gaussian waves, so that the retrieved field will carry the transverse phase contained in the writing field $W$. However, to respect Eq. (1) this phase information has to be projected into the plane transverse to the new propagation axis $z^{\prime}$.  Such a description is at  the basis of  the degenerate four-wave mixing process, known to generate phase conjugation of a probe beam \cite{Hellwarth77}.  In our case, the writing beam $W$ is a LG mode with a topological charge $\ell$, $i.e.$, a field with an azimuthal phase distribution $\ell\phi$ around its propagation axis $z$. The reading process by the $R$ beam transfers this phase into the diffracted beam whose propagation direction is not strictly equal to the direction of $W$. However, as long as the angle $\theta$ between the grating writing beams remains small, we expect that  the corresponding OAM per photon $\ell\hbar$  can be transferred to the retrieved beam. This may be viewed as an application of the conservation law of OAM,  $\vec{L}_W +\vec{L}_R= \vec{L}_{W^{\prime}}+\vec{L}_C$. Such a law, which would have to be vectorial, and cannot be scalar as in \cite{Guo06}, requires at minimum that there are no depolarizing processes \cite{Bloch81} and that finally the exchange of OAM between light and atoms, which is usually distributed between internal and external momentum, is recovered in the nonlinear optical process. On the opposite, for large separation angles one can show that, in the limiting case of $\theta=90^{o}$, no transfer of OAM axis can be performed in the FWM: indeed, in this case symmetry arguments, similar to those previously used for polarization in FWM processes \cite{Bloch81}, can be applied. 

In order to corroborate the above qualitative analysis, we have performed a mode decomposition to analyze the distribution of OAM components of the retrieved beam. To do that, we consider an incident pure LG mode with topological charge $\ell$ and radial mode index $p=0$, so described in the cylindrical coordinates $(\rho, \phi, z)$, at the plane $z=0$, as

\begin{equation}
{\cal E}_W(\vec{r})=LG_{p=0}^{\ell}(\rho,\phi)={\cal E}_0(\frac{\rho \sqrt2}{ w_0})^{|\ell|}e^{-\frac{\rho^2}{ w_0}}e^{i\ell\phi},
\end{equation}
where $w_0$ and ${\cal E}_0$ are, respectively, the beam waist and amplitude. According to our experimental situation, we consider only single annular LG modes with $p=0$ for $W$ beam and assume  the incident writing beam $W^{\prime}$ as being a gaussian mode and the reading beam $R$ as a nearly plane wave. Eqs. (1) and (2) are then used to determine the retrieved field mode as a superposition of $LG_{p^\prime=0}^{\ell^\prime}(\rho^\prime,\phi^\prime)$ modes, defined in terms of a new cylindrical coordinate system $(\rho^{\prime}, \phi^{\prime}, z^{\prime}$), obtained by rotating the original coordinate system by an angle $\theta$ around the $x$ axis, which is orthogonal to both $z$ and $z^{\prime}$ axes defined in Fig. 1. The two cylindrical coordinates are simply related and using the orthogonality relation for LG modes associated with different azimuthal indices, we have numerically calculated, at the plane $z^{\prime}=0$, the corresponding decomposition for different values of the incident topological charge $\ell$ and different angles $\theta$. In Fig. 4 we show the obtained decomposition for four incident LG modes ($\ell =0, 1, 2, 3$) for four angles values. We clearly see that for a reasonable range of angles, $i.e.$, for $\theta \leq 10^o$  the main component of OAM of the retrieved beam is $\ell^{\prime}=\ell$. As $\ell$ increases other $\ell^{\prime}$ values appear in the OAM-spectrum of the retrieved beam. It is worth noticing that the OAM-spectrum only contains $\ell^{\prime} \geq\ell$ values and with same parity. At angle $\theta=2^o$, value used in the experiment, the  contribution of the OAM components different from $\ell$, mostly $\ell^{\prime}=\ell+2$ mode, is less than $0.1\%$.  However, at larger $\theta$ angles the OAM components with $\ell^{\prime}>\ell$ strongly contribute with comparable or even higher amplitudes. Therefore, the conservation law of OAM within the field modes breaks down, as qualitatively  discussed above.
\begin{figure}[htb]
  \centerline{\includegraphics[width= 9.0cm, angle=0]{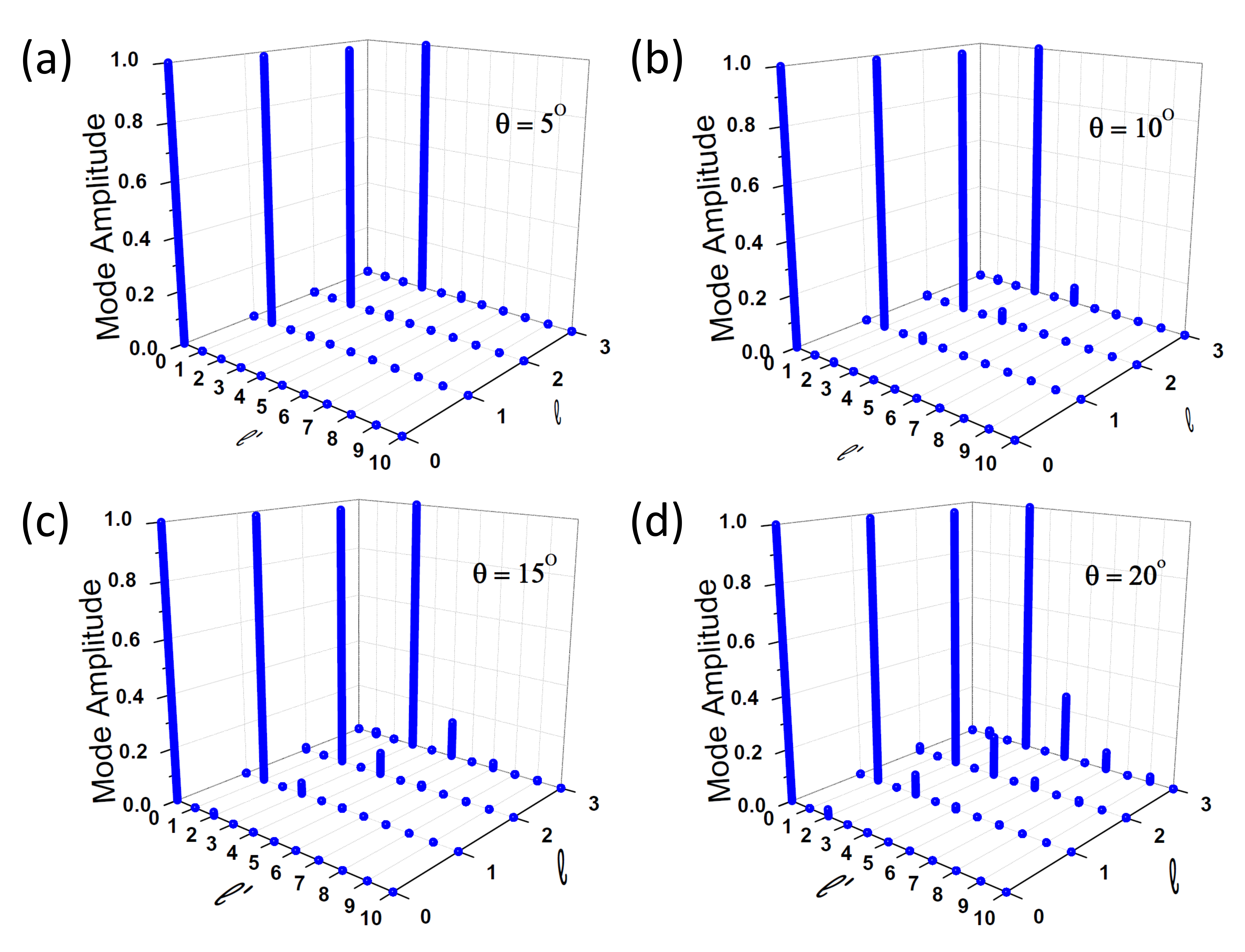}}
  \vspace{-0.2cm}
  \caption{(Color online) Decomposition of the retrieved beam mode in terms of LG modes (azimuthal mode index $\ell^{\prime}$, with $p^\prime=0$) propagating along the $z^{\prime}$ direction, corresponding to a pure incident LG mode (azimuthal mode index $\ell$, with $p=0$) for different angles between the $z$ and $z^{\prime}$ axes: a) $\theta =5^o$,  b) $\theta =10^o$,  c) $\theta =15^o$, and  d) $\theta =20^o$ . For each value of $\ell$, the mode amplitudes are normalized by its maximum value. In the calculation, the ratio between the waists of beams $W^\prime$ and $W$ is 1.4, consistent with the experimental value.}
  \label{fig:Fig4}
\end{figure}

 \section{Conclusion}
 
\indent In summary, we have experimentally demonstrated that light OAM can be stored in an atomic ensemble and retrieved along a nearly different direction. Although the experimental demonstration was restricted to small angle, a mode decomposition calculation shows that the OAM off-axis transfer would work for angles as large as $10^o$. Off-axis  retrieval were also demonstrated for superpositions of LG beams with a gaussian beam.  We have also observed Larmor oscillations in this non-collinear retrieval of OAM demonstrating a simple manipulation and retrieval of OAM along a different axis. We believe that our demonstration of the exchange of OAM in different directions could bring new insights in quantum information processes involving OAM photons, including the possibility of manipulating and distributing OAM-encoded information, with an user controlled time, in different directions in space.

\begin{acknowledgement}
We acknowledge S. S. Vianna and A. A. Monteiro for helping us with the SLM for computational assistance.
This work was supported by the Brazilian agencies CNPq, FACEPE,
INCT/CNP. We also thank CAPES-COFECUB (Ph 740-12) for the support of Brazil-France cooperation.
\end{acknowledgement}

\indent\emph{Permanent addresses}

\emph{($\star$) Unidade Acadmica do Cabo de Santo Agostinho, Universidade Federal Rural
de Pernambuco, BR 101 SUL, Km 97 - S/N, Cabo de Santo Agostinho,
PE, Brazil.}

\emph{$(\ast)$ Instituto de F'sica, Facultad de Ingenier'a, Universidad de la Repœblica, J. Herrera y Reissig 565, 11300 Montevideo, Uruguay}


\begin{thebibliography}{99}


\bibitem{Allen04}
M. Padgett, J. Courtial, and L. Allen,  Physics Today
{\bf 57}, 35 (2004).
\bibitem{Torner13}
Juan P. Torres and Lluis Torner, Twisted Photons: Application of Light with Orbital Angular Momentum (Wiley-VCH, 2011).

\bibitem{Allen92}
L.~Allen, M.~W.~Beijersbergen, R.~J.~C.~Spreeuw, J.~P.~Woerdman,
Phys. Rev. A {\bf 45} , 8185  (1992) .

\bibitem{Bechmann-Pasquinucci00}
H.~Bechmann-Pasquinucci, W. Tittel, Phys. Rev. A {\bf 61}, 62308, (2000).

\bibitem{Petrov99}
J.~W.~R. Tabosa, and D.~V. ~Petrov, Phys. Rev. Lett. {\bf 83}, 4967  (1999) .

\bibitem{Barreiro03}
S. Barreiro and J.~W.~R. Tabosa, Phys. Rev. Lett. {\bf 90}, 133001 (2003).

\bibitem{Guo06}
Wei Jiang, Qun-Feng Chen, Yong-Sheng Zhang, Guang-Can Guo, Phys. Rev. A {\bf 74}, 043811  (2006) .

\bibitem{Andersen06}
M.~F.~Andersen, C.~Ryu, P.~Clade, V.Natarajan, A.~ Varizi, K.~Helmerson, and
W.~D.~Phillips, Phys. Rev. Lett. {\bf 97},170406 (2006) .

\bibitem{Pugatch07}
R. Pugatch, M. Shuker, O. Firstenberg, A. Ron, and N. Davidson,
Phys. Rev. Lett. {\bf 98}, 203601 (2007) .

\bibitem{Moretti09}
D. Moretti, D. Felinto, and J. W. R. Tabosa,
Phys. Rev. A {\bf 79}, 023825, (2009).

\bibitem{Dong13}
Dong-Sheng Ding, Zhi-Yuan Zhou, Bao-Sen Shi, and Guang-Can Guo
Nat. Commun. {\bf 4}, 2527 (2013).

\bibitem{Nicolas14}
A.~Nicolas, L.~Veissier, L.~Giner, E.~Giacobino, D.~Maxein, and J.~Laurat
Nat. Photonics {\bf 8}, 234, (2014).


\bibitem{Marino08}
A.~M.~Marino, V.~Boyer, R.~C.~Pooser, P.~D.~ Lett, K.~Lemons and
K.~M.~Jones, Phys. Rev. Lett. {\bf 101}, 093602 (2008) .


\bibitem{Kozuma06}
R. Inoue, N. Kannai, T. Yonehara, Y. Miyamoto, M. Koashi, and M. Kozuma, 
Phys. Rev. A {\bf 74}, 053809  (2006).


\bibitem{Guo08}
Qun-Feng Chen, Bao-Sen Shi, Yong-Sheng Zhang, Guang-Can Guo, Phys. Rev. A {\bf 78}, 053810  (2008).

\bibitem{Wu13}

Jinghui Wu, Yang Liu, Dong-Sheng Ding, Zhi-Yuan Zhou, Bao-Sen Shi, and Guang-Can Guo, Phys. Rev. A {\bf 87}, 013845 (2013) .


\bibitem{Molina-Terriza03}

Gabriel Molina-Terriza, Juan P. Torres, and Lluis Torner, Opt. Commun.  {\bf 228}, 155 (2003).

\bibitem{Torres04}

Juan P. Torres, Clara I. Osorio, and Lluis Torner, Opt. Lett.  {\bf 29}, 1939  (2004).

\bibitem{Moretti08}
D.~Moretti, N.~Gonzalez, D.~Felinto, and J.~W.~R. Tabosa, Phys. Rev. A {\bf 78}, 023811 (2008) .


\bibitem{Hellwarth77}
R. W. Hellwarth, JOSA {\bf 67}, 1 (1977) .

\bibitem{Bloch81}
D. Bloch and M Ducloy, J Phys B {\bf 14}, 471 (1981).

\bibitem{Singh12}

Pravin Vaity, and R.~P.~Singh, Opt. Lett.  {\bf 37}, 1301 (2012).

\bibitem{Lezama10}
D. Moretti, D. Felinto, J. W. R. Tabosa, and A. Lezama, J. Phys. B: At. Mol. Opt. Phys. {\bf 43}, 115502  (2010).



\end{thebibliography}
\end{document}